\documentstyle[preprint,aps]{revtex}
\begin{document}
%
%
\title{
The Interference Effect of Top Quark Polarization at Hadronic Colliders}
\author{
Darwin Chang$^{(1,2)}$,
Shih-Chang Lee$^{(2)}$ and
Paul Turcotte$^{(1)}$ }
\address{
$^{(1)}$Physics Department,
National Tsing-Hua University, Hsinchu 30043, Taiwan, R.O.C.}
\address{
$^{(2)}$Institute of Physics, Academia Sinica, Taipei, R.O.C.}
%
\maketitle

\begin{abstract}
We derive a simple analytic expression for $q \bar{q}, g g \rightarrow
t \bar{t} \rightarrow b W^+ \bar{b} W^- \rightarrow b \bar{l} \nu_l
\bar{b} l' \bar{\nu_{l'}}$ for on shell intermediate states with the
interference effects due to the polarizations of the $t$ and $\bar{t}$.
We then investigate how this effect may be measured at Tevatron or other
hadronic colliders.
\end{abstract}


The experimental groups at Tevatron has already provided strong evidence
for the top quark with mass around 174 GeV \cite{cdf}.
In the near future, the data
related to the production of the top quark is going to accumulate very
quickly.  Top quark decays very rapidly after its production.  Therefore,
it is challenging to investigate how one can determine the various
detail properties of top quark in the complex hadronic collider environment.

In this paper, we investigate an interesting physical consequence of the fact
that the top quarks that are produced and decayed are supposed to be spin
$1/2$ particles.
Due to the confinement phenomena of the strong interaction, the spin of a
quark is in general very difficult to measure directly.  Even if the
quark is produced in a particular polarized state, the spin information
is quickly averaged out by the hadronization process.
One of the important effect of the polarizations of
unstable particles are that the different polarized intermediate states
can interfere.  For lighter quarks, this effect may be smeared heavily by
the hadronization process.  However, in the case of the top quark, because
of its mass, the decay rate is much faster than for the other quarks.
The top quark decays before the hadronization effect has the time to
smear its polarization information at production\cite{bigi}.  This
allows us to neglect the effect of hadronization.
The same can not be said of any other quarks.
Therefore, in this sense, this polarization effect also provide a unique
possibility of direct observation of the spin of a quark.

We use an analytic helicity technique
developed in ref. \cite{leesc} to calculate the cross section
$q \bar{q}$ and $g g \rightarrow
t \bar{t} \rightarrow b W^+ \bar{b} W^- \rightarrow b \bar{l} \bar{\nu_l}
\bar{b} l' \nu_{l'}$.
This cross section includes 6 particles in the
final state.
We derive simple analytic expressions including the
interference effect, and then give the exact differential cross section
assuming that the top quarks and the $W$ bosons are on-shell.
The decays of $W$ bosons and top quarks are taken into account in the narrow
width approximation. It is clear that the contribution due to the off-shell
top quarks or the off-shell $W$ bosons are negligible.
The interference effects discussed here was also considered before in
ref. \cite{kane}, however, it was studied only numerically and was done in a
rougher approximation.  The analytic expressions obtained here also allows
us to make simple analysis about how such effect may be observable at
Tevatron and future colliders.

To calculate the cross section using an analytic helicity
technique, we split the process
into many subprocesses that can be calculated independently.
In our case, the polarization density matrix is split into two main
sections and can be written as

\begin{eqnarray}
{\it P} &=& {1\over N_{s}} \sum_{
\lambda_1, \lambda_2, \lambda_3, \lambda_3', \lambda_4, \lambda_4'}
{\it P}(i_{\lambda_1\lambda_1}\bar{i}_{\lambda_2\lambda_2}\rightarrow
t_{\lambda_3\lambda_3'}\bar{t}_{\lambda_4\lambda_4'})
\left | \Pi_{t}(r_1) \right |^2
\left | \Pi_{\bar{t}}(r_2) \right |^2
\nonumber\\&&\qquad\qquad\qquad\qquad\times
{\it P}(t_{\lambda_3\lambda_3'}\bar{t}_{\lambda_4\lambda_4'}\rightarrow
b\bar{b}l^+l^-\nu\bar{\nu}) .
\end{eqnarray}
Here the initial the final polarizations are summed over.
$N_{s}$ is the number of initial states to be averaged over.
$N_{s} = 4 (N^2 - 1)^2$ for $gg$ initial states; and
$N_{s} = 4 N^2$ for $q\bar{q}$ initial states.
$\Pi_{q}(r_i)=-i(r_i^2-m_{q}^2)^{-1}$ is the propagator for a scalar
particle. Here, $\Pi_{t}(r_1)$ and $\Pi_{\bar{t}}(r_2)$
represent the polarization independent components of the top quark
propagators with the four-momenta $r_1$ and $r_2$ respectively.
The helicity informations are included in the remaining density matrices.
The polarization density matrix $\it P$ is defined by
\begin{eqnarray}
{\it P}( a_{ij}\ldots \rightarrow b_{kl}\ldots ) &=&
M( a_i\ldots \rightarrow b_k\ldots )
M^\dagger( a_j\ldots \rightarrow b_l\ldots )
\end{eqnarray}
with $M$ being the amplitude for the process.
When no subscript is included for a given particle with spin in $P$,
it implies that its helicity is summed over.
The polarization density matrix
$P(i_{\lambda_1\lambda_1}\bar{i}_{\lambda_2\lambda_2}\rightarrow
t_{\lambda_3\lambda_3'}\bar{t}_{\lambda_4\lambda_4'})$
represents the production of the top quark pair via the processes
$qq\rightarrow t\bar{t}$ or $gg\rightarrow t\bar{t}$.
They have been calculated analytically in detail in ref.
\cite{leesc}.
$P(t_{\lambda_3\lambda_3'}\bar{t}_{\lambda_4\lambda_4'}
\rightarrow b\bar{b}ll\nu\nu)$ is the polarization density matrix
for the decay of the top quark pair into $b$ quarks and leptons.
This density matrix can be further splitted into the product of
the decay density matrix of
$t \bar{t}$ into $W^+ W^-$ bosons and $b \bar{b}$ pairs, and the decay
density matrix of $W^+ W^-$ boson pair into a pair of fermions each.
The main advantage of this technique is that helicity density matrix of each
subprocess can be calculated separately independent of each other.
The results can then be applied to the calculation of different
processes by gluing the density matrix of the subprocesses together.

Therefore the polarized density matrix of the process
$t_{\lambda_3\lambda_3'}\bar{t}_{\lambda_4\lambda_4'}
\rightarrow b\bar{b}ll\nu\nu$ can be written as
\begin{eqnarray}
P(t_{\lambda_3\lambda_3'}\bar{t}_{\lambda_4\lambda_4'}\rightarrow
b\bar{b}l^+l^-\nu\bar{\nu})
&=& \sum_{\lambda_5, \lambda_5', \lambda_6, \lambda_6'}
P(t_{\lambda_3\lambda_3'}\rightarrow W^+_{\lambda_5\lambda_5'} b)
P(\bar{t}_{\lambda_4\lambda_4'}\rightarrow
W^-_{\lambda_6\lambda_6'} \bar{b})
\left | \Pi_{W}(p_1) \right |^2
\nonumber\\
&&\qquad\qquad \times
\left | \Pi_{W}(p_2) \right |^2
P(W^+_{\lambda_5\lambda_5'} \rightarrow l^+ \bar{\nu})
P(W^-_{\lambda_6\lambda_6'} \rightarrow l^- \nu)
\end{eqnarray}
which can be explicitly calculated to be
\begin{eqnarray}
P(t_{\lambda_3\lambda_3'}\rightarrow W^+_{\lambda_5 \lambda_5'} b)
&=& m_t
\left ( { e^2 \left | V_{tb}\right |^2 \over 2 \sin^2\theta_w } \right )
\left [ q_1^{\mu} g^{\nu\rho} - g^{\mu\nu} q_1^{\rho}+
q_1^{\nu}g^{\mu\rho} - i\epsilon^{\sigma\mu\nu\rho}q_{1\sigma}
\right ] \nonumber\\
&&\qquad\qquad\times
\epsilon^*_{1\mu}(\lambda_5)
\epsilon_{1\nu}(\lambda_5')
e^a_{\rho}(r_1) \sigma_{a(\lambda_3 \lambda_3')},
\end{eqnarray}
\begin{eqnarray}
P(\bar{t}_{\lambda_4\lambda_4'}\rightarrow W^-_{\lambda_6\lambda_6'} \bar{b})
&=& m_t
\left ( { e^2 \left | V_{tb}\right |^2 \over 2 \sin^2\theta_w } \right )
\left [ q_2^{\mu} g^{\nu\rho} - g^{\mu\nu} q_2^{\rho}+
q_2^{\nu}g^{\mu\rho} + i\epsilon^{\sigma\mu\nu\rho}q_{2\sigma}
\right ] \nonumber\\
&&\qquad\qquad\times
\epsilon^*_{2\mu}(\lambda_6)\epsilon_{2\nu}(\lambda_6')
e^a_{\rho}(r_2) \bar{\sigma}_{a(\lambda_4 \lambda_4')},
\end{eqnarray}
\begin{eqnarray}
P(W^+_{\lambda_5\lambda_5'}
\rightarrow l^+ \nu)
&=& \left ( {e^2\over \sin^2\theta_w} \right )
\left [
k_{1}^{\mu} l_{1}^{\nu}+k_{1}^{\nu} l_{1}^{\mu}
-g^{\mu\nu}k_1\cdot l_1
+ i\epsilon^{\sigma\mu\rho\nu}
k_{1\sigma}l_{1\rho}
\right ]
\epsilon_{1\mu}(\lambda_5)\epsilon_{1\nu}^{*} (\lambda_5')
\end{eqnarray}
and
\begin{eqnarray}
P(W^-_{\lambda_6\lambda_6'}\rightarrow l^- \bar{\nu})
&=& \left ( {e^2\over \sin^2\theta_w} \right )
\left [
k_{2}^{\mu} l_{2}^{\nu}+k_{2}^{\nu} l_{2}^{\mu}
- g^{\mu\nu}k_2\cdot l_2
- i\epsilon^{\sigma\mu\rho\nu}
k_{2\sigma}l_{2\rho}
\right ]
\epsilon_{2\mu}(\lambda_6)\epsilon_{2\nu}^{*} (\lambda_6').
\end{eqnarray}
These results include summations over the helicities of the final
state fermions.
The four vectors $k_i$, $l_i$ and $q_i$ correspond to the momenta of
the neutrinos, the charged leptons and the b quarks respectively.
The subscript 1 is used to label the momenta related to the decay
products of the top quark, while the subscript 2 is used for those
related to the decay of the anti-top quark.
$\epsilon_{1\mu}(\lambda_5)$
represents the polarization vectors of $W^+$ with helicity
$\lambda_5$, while
$\epsilon_{2\nu}(\lambda_6)$
represents those of the $W^-$ with helicity
$\lambda_6$.
In our notation, the information about the top helicity is carried by
the Pauli matrix $\sigma_i$ and $\sigma_0 (= 1)$
with component $(\lambda_3\lambda_3')$.
For the anti-top, we used the conjugate matrix
$\bar{\sigma}_a = \sigma_2 \sigma_a \sigma_2 $
with component $(\lambda_4\lambda_4')$.

The tensor $e^a (r_i)$, $(a = 0,1,2,3)$, were introduced in
ref.\cite{leesc} to evaluate the product of fermion wave functions,
$u(r,\lambda)\bar{u}(r,\lambda')$,
in the amplitude squared for the cross section.  They are defined as
\begin{eqnarray}
u(r,\lambda)\bar{u}(r,\lambda')&=& {1\over 2} (m+p\!\!\! \slash )
\left ( e\!\!\! /^0 (r) \sigma_{0(\lambda\lambda')} +
 e\!\!\! /^i (r)\gamma_5 \sigma_{i(\lambda\lambda')} \right ),
\end{eqnarray}
\begin{eqnarray}
v(r,\lambda)\bar{v}(r,\lambda')&=& {1\over 2} (m-p\!\!\! \slash )
\left ( e\!\!\! /^0 (r) \bar{\sigma}_{0(\lambda\lambda')} +
 e\!\!\! /^i (r)\gamma_5 \bar{\sigma}_{i(\lambda\lambda')} \right ).
\end{eqnarray}
The explicit form of the tensor $e^a (r_i)$ can be found in the reference
\cite{leesc,hagi}, but
in our case, we only need the following properties,
\begin{eqnarray}
\sum_\mu r^\mu e^a_\mu (r) &=&
m_t \delta^{0a} ,\\
\sum_\mu e^{a\mu}(r) e^b_{\mu}(r) &=& g^{ab},\\
\sum_{k=1}^3 e^k_{\mu}(r) e^k_{\nu}(r)
&=& -g_{\mu\nu} + r_{\mu}r_{\nu}/ m_t^2.
\end{eqnarray}
where $r$ is the 4-momentum of either the top quark or the anti-top quark.
The spatial component, $e^i_{\mu}(p)$ (i.e. $i = 1,2,3$),
were first introduced by Bouchiat and Michel \cite{michel}.

One can then combine the above results for all the subprocesses and
use the narrow width approximation for the $W$ boson propagators to
obtain
\begin{eqnarray}
P(t_{\lambda_3\lambda_3'}\bar{t}_{\lambda_4\lambda_4'}\rightarrow
b\bar{b}ll\nu\nu)
&=& 16 m_t^2 \left (
{ e^4 \left | V_{tb}\right |^2 \over 2 \sin^4\theta_w } \right )^2
(l_{1} \cdot e_{1}^a)(k_{1}\cdot q_{1})
(l_{2} \cdot e_{2}^b)(k_{2}\cdot q_{2})
\sigma_{a(\lambda_3\lambda_3')}
 \otimes \bar{\sigma}_{b(\lambda_4\lambda_4')}
\nonumber\\
&&\times
{\pi^2\delta \left ( p^2_{1} - m^2_W \right )
\delta \left ( p^2_{2} - m^2_W \right )
\over ( M_W \Gamma_W )^2 }.
\end{eqnarray}

With this, one can then proceed to glue together the density matrix for
the production, which has been given in Ref.\cite{leesc}, and those
for the decays of the top quarks.
For the top and anti-top propagators, we shall use also
the narrow width approximation.
These provide another two delta functions.
After that, we can write the phase space for 6 final state particles
in a such way that it will be straight forward
to integrate out all the delta  function.
The phase space can be rewritten from the standard way \cite{phase},
\begin{eqnarray}
d_6 ( T \rightarrow k_{1} l_{1} q_{1} k_{2} l_{2} q_{2} ) &=&
(2\pi)^4 \delta^4 ( T - k_{1} - l_{1} - q_{1} - k_{2} - l_{2} - q_{2})
\nonumber\\
&&
{d^3l_{1}\over (2\pi)^3 2 E_{l_{1}}}
{d^3k_{1}\over (2\pi)^3 2 E_{k_{1}}}
{d^3q_{1}\over (2\pi)^3 2 E_{q_{1}}}
{d^3l_{2}\over (2\pi)^3 2 E_{l_{2}}}
{d^3k_{2}\over (2\pi)^3 2 E_{k_{2}}}
{d^3q_{2}\over (2\pi)^3 2 E_{q_{2}}}\\
\nonumber\\
&=& {1\over (2\pi)^4}
d_2 (T\rightarrow r_{1} r_{2})
d_2 (r_{1} \rightarrow p_{1} q_{1})
d_2 (r_{2} \rightarrow p_{2}  q_{2})\nonumber\\
&&\times
d_2 (p_{1} \rightarrow l_{1} k_{1})
d_2 (p_{2} \rightarrow l_{2} k_{2})
d r_1^2 d r_2^2 d p_1^2 d p_2^2
\end{eqnarray}
where
$T^\mu = i^\mu_1+i^\mu_2$,
with $i_1$ and $i_2$ defined as the four-momenta of the initial partons
and
\begin{eqnarray}
d_2 ( a \rightarrow b c) &=&
(2\pi)^4 \delta^4 ( a-b-c)
{d^3b\over (2\pi)^3 2 E_{b}}
{d^3c\over (2\pi)^3 2 E_{c}}
\end{eqnarray}
is the phase space for a particle with four-momentum $a$
decaying into particles with four-momenta $b$ and $c$.

Combining the above results, we obtain directly the
differential cross section
\begin{eqnarray}
d\sigma_{ii} &=& {P \over 4 i_1\cdot i_2}
d_6 ( T \rightarrow k_{1} l_{1} q_{1} k_{2} l_{2} q_{2} )
\nonumber\\
&=&
{4 \pi^2 \alpha_s^2\over N} \left (
{ e^4 \left | V_{tb}\right |^2 \over 2 \sin^4\theta_w } \right )^2
(k_{1} \cdot q_{1})(k_{2}\cdot q_{2}) l_{1\alpha} l_{2\beta}
I_{ii}^{\alpha\beta} \,
{d_2 (T\rightarrow r_{1} r_{2})\over i_{1}\cdot i_{2}}
\nonumber\\
&&\times
{d_2 (r_{1} \rightarrow p_{1} q_{1})
d_2 (r_{2} \rightarrow p_{2}  q_{2})
d_2 (p_{1} \rightarrow l_{1} k_{1})
d_2 (p_{2} \rightarrow l_{2} k_{2})\over
\left ( M_W \Gamma_W M_t \Gamma_t \right )^2 },
\end{eqnarray}
$N$ is the number of color and the indices
$ii$ designate the initial particles.
We have the result for the
$g g \rightarrow t \bar{t}$ case
\begin{eqnarray}
I_{gg}^{\alpha\beta}
 &=& \left ( x - 1 - {1\over N^2-1} \right ) \left \{
2  r_{1}^\alpha r_{2}^\beta  \left [ {-1\over x}
+ {1\over \gamma^2} \left ( 1- {x\over 2\gamma^2}\right )
+ 1 - {x\over 2\gamma^2} \right ] \right . \nonumber\\
&&
-{1\over 2\gamma^2}\left( 1 - {x\over \gamma^2} \right )
\left (  r_{1}^\alpha T^\beta  +
 T^\alpha r_{2}^\beta  \right )
+ m^2_t \left ( {-1\over x} + {1\over \gamma^2}
\left ( 1 - {x\over 2\gamma^2} \right )
 \right )   g^{\alpha\beta} \nonumber\\
&&  \left .
+ {1\over 4\gamma^2}
\left ( 1 - {x\over 2 \gamma^2} \right )
\left [ T^\alpha T^\beta
- R^\alpha R^\beta
-\left ( {u-t\over s} \right ) \left (
 R^\alpha T^\beta - T^\alpha R^\beta  \right ) \right ]
\right \}
\end{eqnarray}
and for the $q \bar{q}\rightarrow t \bar{t}$ case
\begin{eqnarray}
I_{qq}^{\alpha\beta}
 &=& \left ( {N^2-1 \over 2 N} \right ) \left \{
2 r_{1}^\alpha r_{2}^\beta  \left [ 1
-{1\over x} \left (1 - {x\over 2\gamma^2} \right )\right ] \right .
\nonumber\\
&&
-{1\over 2\gamma^2}\left ( r_{1}^\alpha T^\beta  +
 T^\alpha r_{2}^\beta \right )
- {m^2_t\over x} \left ( 1 - {x\over 2 \gamma^2} \right )
 g^{\alpha\beta}\nonumber\\
&& \left .
+ {1\over 4\gamma^2}
\left [
 T^\alpha T^\beta
- R^\alpha R^\beta
-\left ( {u-t\over s} \right ) \left (  R^\alpha T^\beta
- T^\alpha R^\beta \right ) \right ]
\right \}.
\end{eqnarray}
The $s$, $t$ and $u$ are the Mandelstam variable,
$T^\mu = i^\mu_1+i^\mu_2$,
$R^\mu = i^\mu_1-i^\mu_2$,
$\gamma^2 = s/4m_t^2$ and
$x=s^2/2(t-m_t^2)(u-m_t^2)$.

We want to show the spin effect of the quark. We need to compare
to the case where the top quark is taken to be a spinless particle.
This will allow us to pinpoint the spin effect.
One way to get formula for the spinless case is to take
the average of the top quark helicity both in the
production density matrix and in the decay density matrix independently
before we join them together.
In the usual Monte Carlo simulation for the
experiments, this is indeed the approximation taken by the program.
The cross section can be obtained by substituting
$I_{gg}$ and $I_{qq}$ with
\begin{eqnarray}
I_{\overline{gg}}^{\alpha\beta}
&=& \left ( x - 1 - {1\over N^2-1} \right ) \left \{
  r_{1}^\alpha r_{2}^\beta  \left [ 1 - {1\over x}
+ {1\over \gamma^2} \left ( 1- {x\over 2\gamma^2}\right )
\right ] \right \}
\end{eqnarray}
and
\begin{eqnarray}
I_{\overline{qq}}^{\alpha\beta}
 &=& \left ( {N^2-1 \over 2 N} \right ) \left \{
 r_{1}^\alpha r_{2}^\beta  \left ( 1
-{1\over x} \left (1 - {x\over 2\gamma^2} \right )\right )
\right \}.
\end{eqnarray}
For the last two equation, we should emphasize that
the cross section without the spin correlation for the
top quark is the one that has been used in the literature and by most of
the experimental event simulators.

If one does not observe any correlation between the observables
associated with the final decay products of top decays, the differential
cross sections
can be integrated easily in a covariant notation using well known
relation for integration of two particles phase-space.
One can obtain a very simple expression for the differential cross
section
\begin{eqnarray}
\label{diff-cross}
{d\sigma_{ii}\over d\cos\theta} &=&
{1\over 9\times 2^{19} \pi^3} { \alpha_s^2\over N} \left (
{ e^4 \left | V_{tb}\right | \over 2 \sin^4\theta_w } \right )^2
\left ( { M_W\over \Gamma_W \Gamma_t m_t } \right )^2
\left ( 1- {1\over\gamma^2} \right )^{1/2}
\lambda(1,y_b,y_W) \nonumber\\
&&\times\left [ 1+y_b-2y_W + {(1-y_b)^2\over y_W} \right ]^2
{r_{1\alpha} r_{2\beta}\over s} I^{\alpha\beta}_{ii}
\end{eqnarray}
where $\theta$ is
the angle between the incoming parton $i_{1}$
and the top quark $r_{1}$, $y_b = m_b^2/m_t^2$,
$y_W = m_W^2/m_t^2$ and
$\lambda(a,b,c)= a^2+b^2+c^2 -2 ab -2ac -2bc$.
The integration over one of the top branch gives result
proportional to $r_1^\alpha I_{\alpha\beta}$ or $r_2^\beta
I_{\alpha\beta}$.
As expected, after the integration of final phase space variables,
spinless approximation gives the same result as the exact one.
This can be seen as a consequence of the identities
\begin{eqnarray}
r_{1\alpha}
I_{gg}^{\alpha\beta}
&=& r_{1 \alpha}
I_{\overline{gg}}^{\alpha\beta},\quad
 r_{2 \beta}
I_{gg}^{\alpha\beta}
= r_{2 \beta}
I_{\overline{gg}}^{\alpha\beta}
\end{eqnarray}
and
\begin{eqnarray}
r_{1\alpha}
I_{qq}^{\alpha\beta}
&=& r_{1 \alpha}
I_{\overline{qq}}^{\alpha\beta},\quad
r_{2 \beta}
I_{qq}^{\alpha\beta}
= r_{2 \beta}
I_{\overline{qq}}^{\alpha\beta}.
\end{eqnarray}
The result not only confirms that the interference effects indeed vanish
when integrated over the complete phase space, but also,
that if one integrates over all the phase space
variables associated to the decay products of just $\it{one}$ of the top
quarks, the spin interference effect will disappear.
Equivalently if we measure any observables related to only one of the
top(or anti-top) quark, the effect will not show up.
Therefore, to measure interference effect, we need to find observables
that correlate the kinematic variables associated with the particles from
the top quark decay and those from the anti-top quark decay.
Intuitively, it is not obvious what is the best correlated observable
which can probe this interference effect.  For the rest of this paper, we
shall present our attempts in this direction.

Before we discuss how to probe the interference effect, one should note
that Eq.(\ref{diff-cross}) can be further simplified.
The widths can be replaced by branching ratios using
$\Gamma_t = \Gamma_{t \rightarrow W b } / Br(t \rightarrow W b)$ and
$\Gamma_W = \Gamma_{W \rightarrow \bar{l} \nu_l} /
Br(W \rightarrow \bar{l} \nu_l)$.  The partial widths can be
calculated trivially at tree level, and the branching ratios can be
evaluated by summing over all the possible
channels for top desintegration.
The $cos\theta$ integration can also be performed to reproduce the
well-known result\cite{phase}
\begin{eqnarray}
\sigma_{qq} = { 4 \pi \alpha_s^2 \over 9 s} \beta
\left ( 1 - {\beta^2 \over 3} \right )
\end{eqnarray}
and
\begin{eqnarray}
\sigma_{gg} = {\pi \alpha_s^2 \over 48 s} \beta
\left [ \left ( 66-36\beta^2+2 \beta^4 \right )
{1\over 2\beta}\ln \left ( {1+\beta\over 1-\beta} \right )
-59 + 31 \beta^2   \right]
\end{eqnarray}
where $\beta^2 = 1 - 1/\gamma^2$.

To probe the interference effect, we shall look for observables
related to the final decay products, instead of using directly the top
momenta which need to be reconstructed.
The easiest observables to try are those related to the two charged
leptons (electron or muon) from the leptonic decays of two $W$'s. The
advantage is that these particles can be easily observed in a
hadron collider.  On the other hand,
the dilepton decays has
lower branching ratios, which reduces the statistics.
For this case, we have investigated the effect of interference on the
distribution of
(1) their total energy $E_{l_{1}}+E_{l_{2}}$, (2) their total z-momentum
$l_{1z}+l_{2z}$, (3) an orthogonal combination $l_{1z}-l_{2z}$,
(4) the cosine of the angle between these two charged leptons and (5) the
asymmetries $A_p$ and $A_{\perp}$ to be defined later.
To get these distributions as a function of the observables,
the numerical integration program (VEGAS) is used. The top quark mass is
taken to be $m_t =$ 176 GeV in all numerical analyses.

In Fig. 1, we show the differential cross section
$d\sigma/dv_1$
for the quark-antiquark collision
where $v_1 = (E_{l_{1}}+ E_{l_{2}} )/\sqrt{s}$. We show curves
a, b, c, d and e for $\sqrt{s} = 400, 500, 700, 1000$ and $2000$ GeV.
At each energy,
we plot the cross section with (without) interference effect with
solid (dash) lines.
We found that,
for the quark-antiquark collision,
in the central (peak) region the interference effect decreases
the cross section
while at the two shoulders, it enhances the cross section.
In addition, while the overall cross section decreases roughly as $1/s$
as $s$ increases, the percentage decrease of the maximum value of
$d\sigma/dv_1$
due to the interference effect actually increases as $s$.
(For example, for $\sqrt{s} = 400$ GeV, the peak of the cross section
decreases
from 1.17 pb to 1.15 pb, while, for $\sqrt{s} = 1$ TeV, it decreases from
.19 pb to .17 pb.)
In other words, comparing to
the spinless case, the effect of the interference gets stronger as the
total energy increases.

In Fig. 2.I  and 2.II, we show the same differential cross section
$d\sigma/dv_1$ for the gluon fusion case.
Again,
we plot the cross section with (without) interference effect using
solid (dash) lines.
In Fig. 2.I, we use $\sqrt{s} = 400, 500$ GeV for curves a, b
and, in Fig. 2.II, we use $\sqrt{s} = 700, 1000$
and $2000$ GeV for curves c, d and e respectively.
For $\sqrt{s} < 700$ GeV, unlike the previous case, the cross section
increases with $s$ as is well-known.
And here, also in stark contrast with the previous case,
the effect of the interference
increases the values of the cross section in the region near the
peak, while making the values at the two shoulders decrease.
However, for higher total energy $\sqrt{s}$,
the behavior become the same as the case for $q\bar{q}$.
One should also note that the interference effects for gluon fusion case
appear to be always smaller than that for the quark-antiquark
production process.

In Fig. 3 and Fig. 4, we show distribution of $d\sigma/dv_2$
as a function of
$v_2 = (l_{1}^z + l_{2}^z)/\sqrt{s}$ for
quark-antiquark ($q\bar{q}$)
production and for gluon fusion (gg) production respectively.
The $l_{1}^z$ is the $z$-component of the momentum, $l_1$, of the
lepton from top decay where the $z$ is defined to be the direction of
parton $i_1$ which is also the beam axis.  Similarly, one defines
$l_2^z$ with the same definition of $z$-axis.
The curves a, b, c and d in Fig. 3 for $q\bar{q}$ case
correspond to $\sqrt{s} = 400, 500, 700$ and $1000$ GeV
while, in Fig. 4 for gg case, the curves a, b and c correspond to
$\sqrt{s} = 400, 450$ and $500$ GeV.
In $q\bar{q}$ case, the interference effect always increases the peak of
the distribution.   In addition, the percentage increase of the effect
decreases with $\sqrt{s}$ but only mildly.
For gluon fusion case, below $\sqrt{s}=700$ GeV, the peak of the
distribution is reduced by the interference effect and then become
larger for $\sqrt{s}$ over 1 TeV.  However, the effect is smaller than
the $q\bar{q}$ case in general.

We have also analyzed the effect of the interference on the
distribution of $d\sigma/dv_3$
as a function of
$v_3 = (l_{1}^z - l_{2}^z)/\sqrt{s}$.
We found the interference effect to be negligibly small
at $\sqrt{s}$ below 700 GeV for $q\bar{q}$ production.  The effect is
even smaller for all the other $\sqrt{s}$ and for the case of gg production.

A potentially very interesting way
to observe the interference effect is to
measure the distribution of angle between the two final leptons.
We get a good signal of the interference effect
near the threshold of the $t\bar{t}$ production.
For $q\bar{q}$ production, we plot in Fig. 5.II, the differential cross
section
$d\sigma/d\cos\theta_{ll}$
as a function of $\cos\theta_{ll}$.
The curves a, b and c correspond
to $\sqrt{s} = 400, 500$ and $700$ GeV and $\theta_{ll}$ is the
angle between the two lepton.
For comparison, we plot in Fig. 5.I, the differential cross section
$d\sigma/d\cos\theta_{WW}$
as a function of $\cos\theta_{WW}$
for the same $\sqrt{s}$, where
$\theta_{WW}$ is the angle between the two W bosons.
Somewhat unexpectedly, we found that the
signal of interference is much stronger in $d\sigma/d\cos\theta_{ll}$
than that in the case of $d\sigma/d\cos\theta_{WW}$.
This difference seems to indicate that at the level of $W^+ W^-$ in
addition to the asymmetry carried by the the kinematic variable
$\cos\theta_{WW}$, some asymmetry due to the interference is carried by
the helicity information of the $W$ bosons.  This helicity-related
asymmetry is turned into the leptonic angular asymmetry as $W$ bosons
decay.  As a result, the angular asymmetry for the angle between the leptons
provides more sensitive
signal than the angular asymmetry for the angle between the $W$ bosons.
We can see that both
side of the distribution for leptons case change by more then 25 \% .
For increasing $\sqrt{s}$ up to 1 TeV, the
$q\bar{q}$ collision still show a large difference between the
cross sections with and without the interference effect.

For the gluon fusion, we also produce the distributions
$d\sigma/d\cos\theta_{WW}$ and
$d\sigma/d\cos\theta_{ll}$, as shown in
Fig. 6.I and Fig. 6.II.
The curves a, b and c correspond to $\sqrt{s} = 400,
450$ and $500$ GeV.  At $\sqrt{s} = 400$, the two distributions
for $d\sigma/d\cos\theta_{ll}$ differ
dramatically from each other up to a factor 2.
However, the difference is reduced quickly as $\sqrt{s}$ increases, the
difference is
negligible at $\sqrt{s}=$ 700 GeV.  The most important characteristic
is that, in contrast with Fig. 1--4, the signal does not just consist of a
change in the width or height, but a
change in the shape of whole distribution and therefore may provide an
easier signature for experiments.

The other interesting observables are the following two new kinds
of asymmetries.  In these asymmetries, one looks at the directions of
momenta of two
particles and compares if they are on the same side of a defining plane or
on the opposite side.
The asymmetry can be defined as
$A =
(N_{\text{same}}-N_{\text{opposite}})/(N_{\text{same}}+N_{\text{opposite}})$.
For the first asymmetry, the plane is the production plane defined
by the beam axis and the direction of the top quark.  The
two incoming partons and the two top quarks lie on this plane.  The
asymmetry associated with this plane will be called $A_p$.
For the second asymmetry, the defining plane is to be the one that is
perpendicular to
the beam axis.  The associated asymmetry will be called $A_{\perp}$.

In Fig. 7a, we show $A_p$ as a function
of $\sqrt{s}$ using the momenta of the two final leptons to define
asymmetry.
As a comparison, in Fig. 7b, we show the same asymmetry but now using
the momenta of the two W bosons as the defining directions.
The solid lines are for the case of $q\bar{q}$ production.  The dotted
lines are for gluon fusion.
The asymmetry should vanish without the interference effect.
Note that just as the earlier cases, the asymmetry is more pronounced
when measured by the final leptons, instead of by the intermediate,
polarized W bosons.  As shown in Fig. 7a, the asymmetry, $A_p$, can be as
large as 5-10 percent.
Experimentally it is also important to note that this asymmetry is invariant
under a boost in the beam direction.  Therefore, as long as the
transverse momenta of the initial partons are small, the asymmetry is
more or less insensitive to the longitudinal momenta of the initial partons.

In Fig. 8, we show $A_{\perp}$ as a function of $\sqrt{s}$.  Here we
present only the result for the case in which the momenta of the two final
leptons are used to define asymmetry.   Just as the previous case, if the
$W$ momenta are used instead of the lepton momenta, the asymmetry is smaller.
The solid line is for $q\bar{q}$ production including the interference
effect.  The dash line is for $q\bar{q}$ production without the
interference effect.  The dotted line is for gg fusion including the
interference effect.  The dash-dot line is for gg fusion without the
interference effect.  As shown in the figure, the asymmetries in all
these channels are all negative with sizes of order .4--.7.
Unfortunately, this asymmetry is sensitive to whether the initial partons
are colliding in their center of mass in the Lab. frame or not.
Therefore, it is probably harder to isolate the effect from the background.

In conclusion, we have obtained an analytic formula for the cross section
of the top quark pair production and their decays
into b quarks and W bosons which, in turns, decay into two fermions.
Our result includes the interference effect due to the
fact that top quarks carry spin and therefore the intermediate
states have more than one spin channels.  A measurement of this
interference effect can be seen as the direct measurement of the spin
effect of the top quarks.
To demonstrate the use of these analytic results, we use them to search
for observables that may provide a good measurement of the interference
effect.  While none of the observables may be straightforward to measure,
we show that appreciable effects might be observable in the
asymmetries defined by the relative angular distributions of the leptons
coming from both W bosons.  More realistic simulation
needs to be carried out before these potential observables can
prove to be useful experimentally.

\acknowledgments

We would like to thank Alexei Sumarokov for useful discussions.
This work is supported
by the National Science Council of Republic of China
under grants
NSC84-2811-M008-001(for S.L),
NSC84-2112-M007-042, NSC84-2112-M007-016(for D.C. and P.T.).

\newpage

\figure{Fig. 1.
The differential cross section $d\sigma/dv_1$
with (without) interference effect using
solid (dash) lines for the quark-antiquark collision.
The curves a, b, c, d and e correspond to
$\sqrt{s} = 400, 500, 700, 1000$ and $2000$ GeV. }

\figure{Fig. 2.
Same as Fig. 1 but for the gluon fusion. }

\figure{Fig. 3.
The differential cross section $d\sigma/dv_2$
with (without) interference effect using
solid (dash) lines for the quark-antiquark collision.
The curves a, b, c and d correspond to
$\sqrt{s} = 400, 500, 700$ and $1000$ GeV. }

\figure{Fig. 4.
The differential cross section $d\sigma/dv_2$
with (without) interference effect using
solid (dash) lines for the gluon fusion.
The curves a, b and c correspond to
$\sqrt{s} = 400, 450$ and $500$ GeV. }

\figure{Fig. 5.
The differential cross section (I) $d\sigma/d\cos\theta_{WW}$
and (II) $d\sigma/d\cos\theta_{ll}$
with (without) interference effect using
solid (dash) lines for the quark-antiquark collision.
The curves a, b and c correspond
to $\sqrt{s} = 400, 500$ and $700$ GeV.}

\figure{Fig. 6.
The differential cross section (I) $d\sigma/d\cos\theta_{WW}$
and (II) $d\sigma/d\cos\theta_{ll}$
with (without) interference effect using
solid (dash) lines for the gluon fusion.
The curves a, b and c correspond
to $\sqrt{s} = 400, 450$ and $500$ GeV.}

\figure{Fig. 7.
The asymmetry $A_p$ using
(a) the two final leptons
and (b) the momenta of the two W bosons
to define directions
The solid (dash) lines correspond to the case of $q\bar{q}$ production
(gluon fusion).}

\figure{Fig. 8.
The asymmetry $A_{\perp}$ as a function of $\sqrt{s}$.
The solid (dash) line is for $q\bar{q}$ production including (without)
the interference effect.
The dotted (dash-dot) line is for gg fusion including (without) the
interference effect.}


\begin{references}
\bibitem{phase} V. Barger and R. J. N. Phillips, {\it Collider Physics},
Addison-Wesley (1987), Appendix B.
Note that our convention for $d_n$ is different from the
one used in the book by $(2\pi)^{4-3n}$.
\end{references}
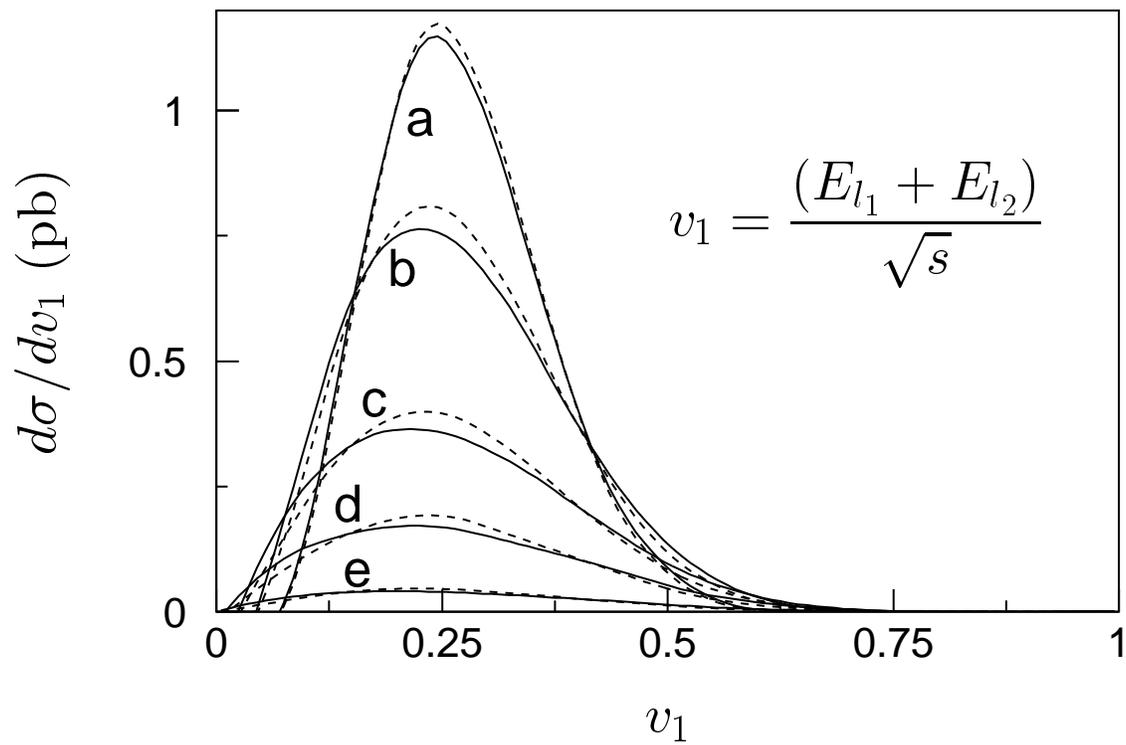
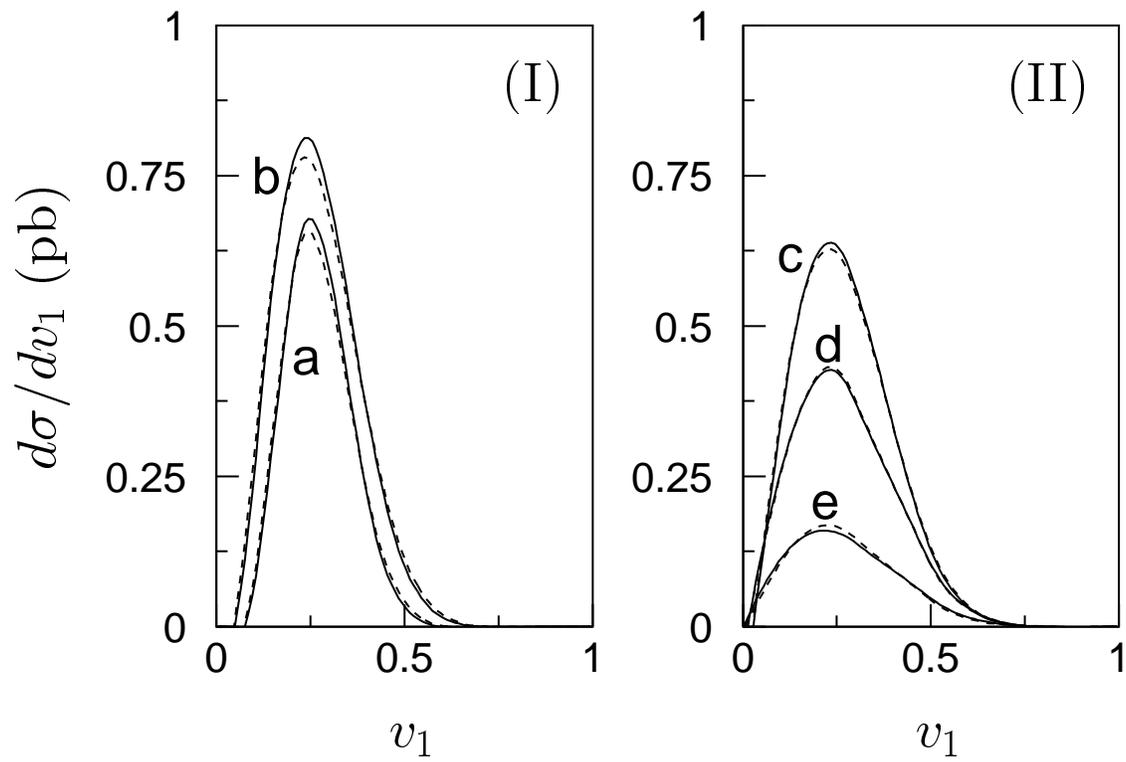
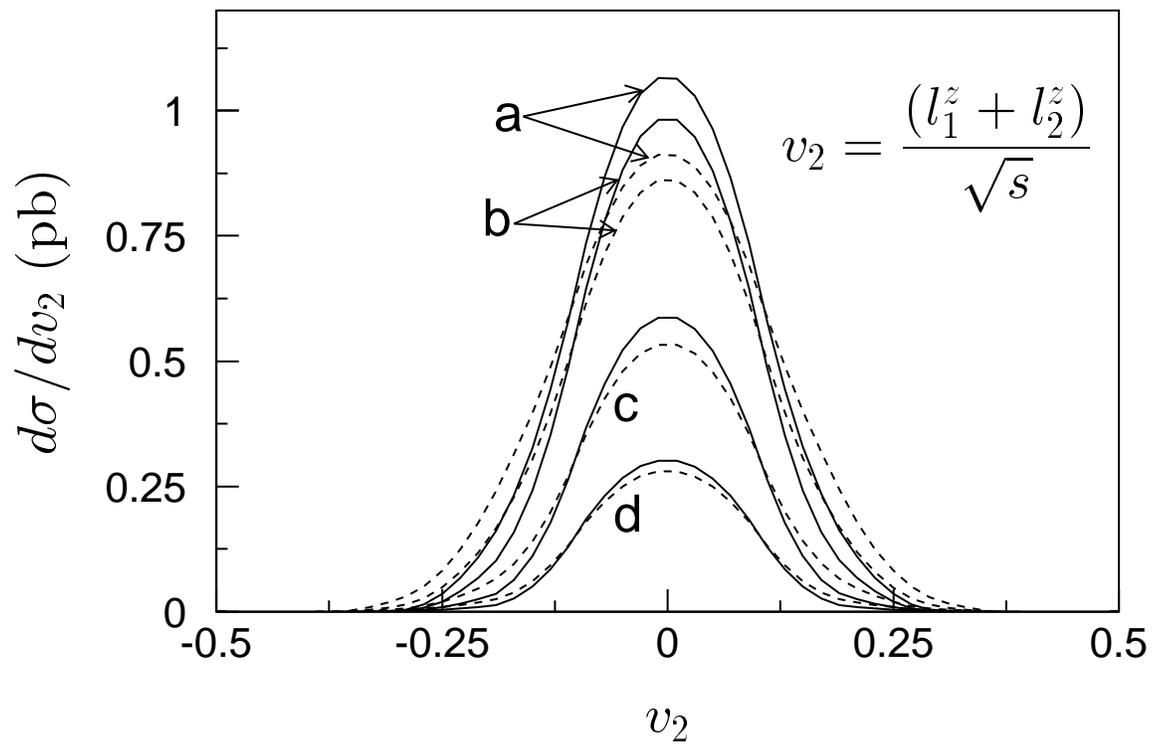
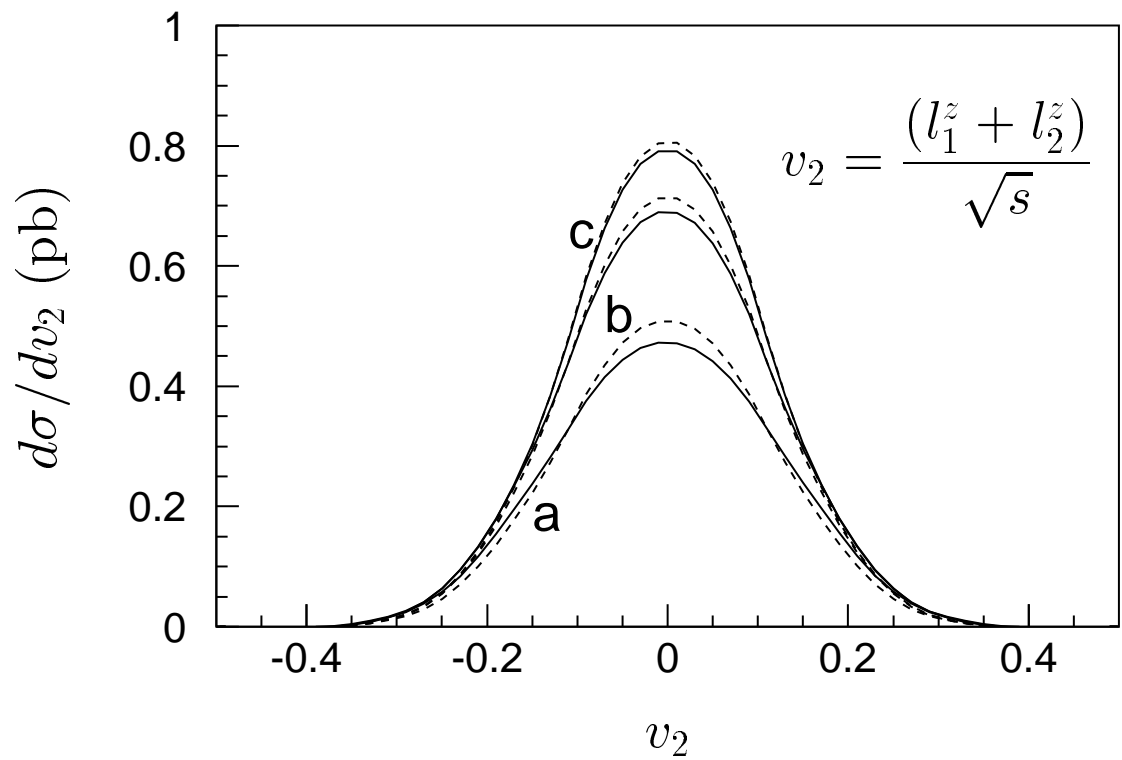
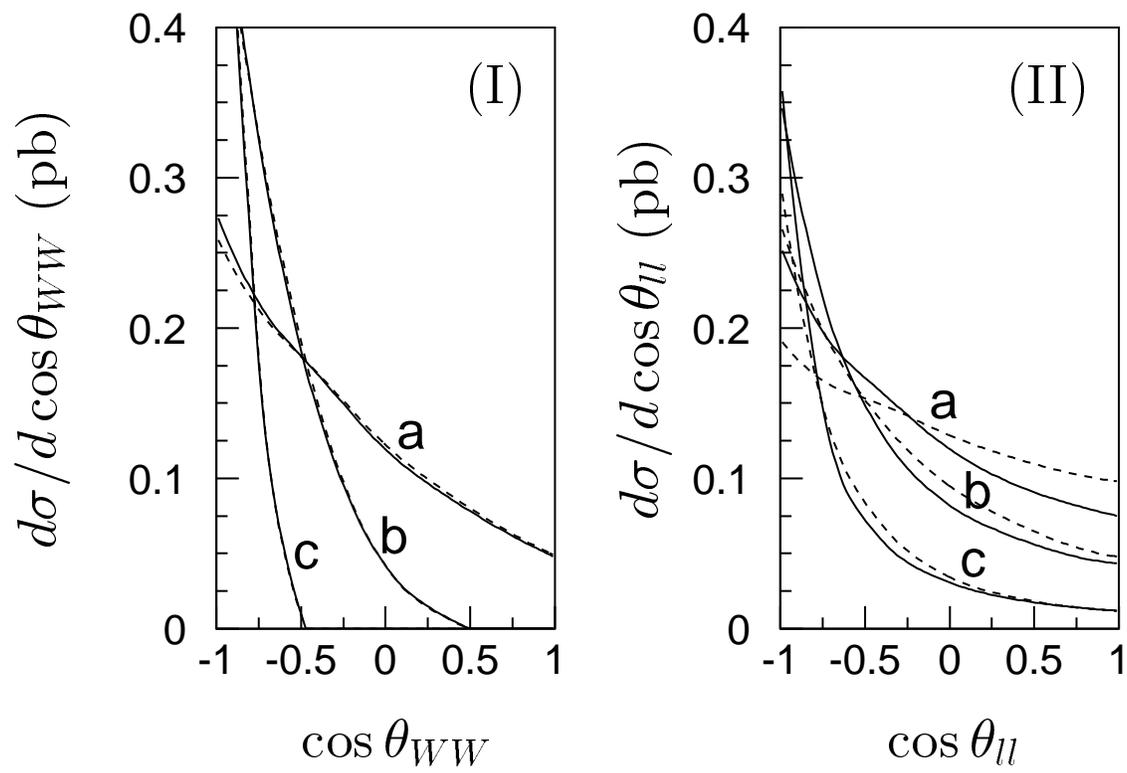
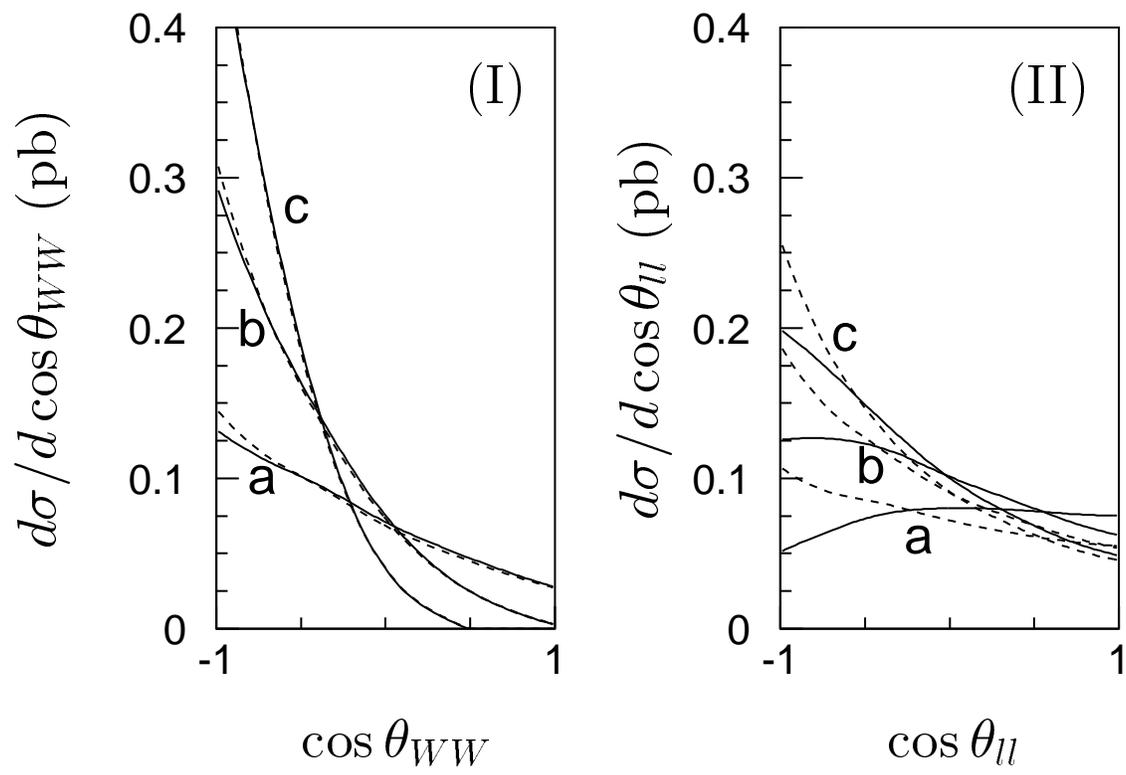
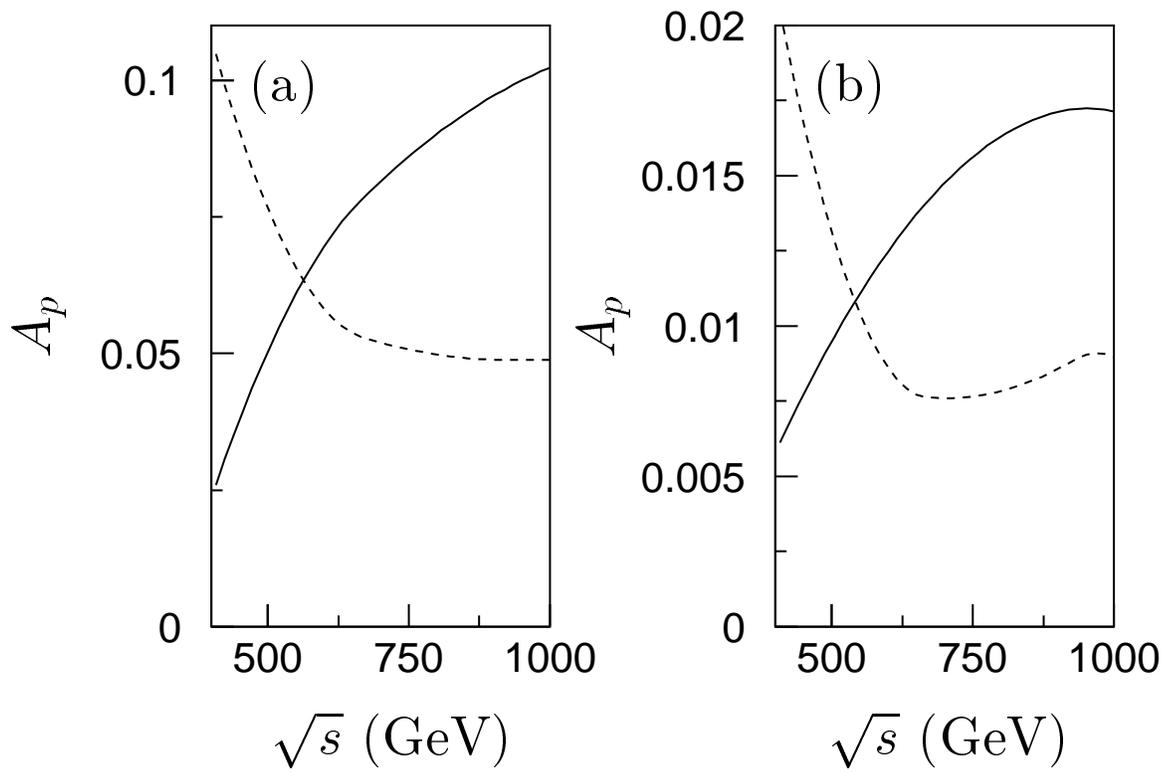
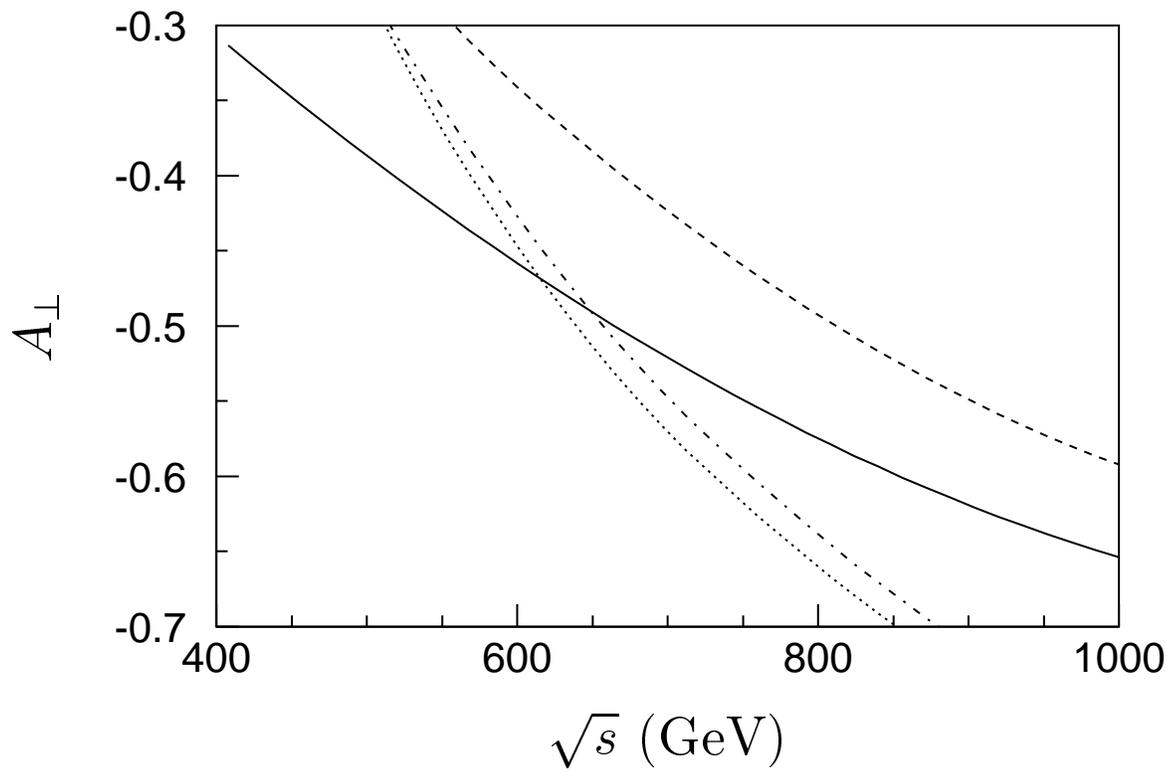
\end{document}